# Channeling defect emission through topological Jackiw-Rebbi states in hBN

**Hugo Quard[1,2,†*], Sergei Nedić[1,2,*], Nika Teran[1,2], Otto Cranwell Schaeper[1,2], Xiaoying Huang [1,2] Shery L. Y. Chang[3] and Igor Aharonovich[1,2,†]**

[1] School of Mathematical and Physical Sciences, University of Technology Sydney, Ultimo, New South Wales 2007, Australia

[2] ARC center of Excellence for Transformative Meta-Optical Systems, University of Technology Sydney, Ultimo, New South Wales 2007, Australia

[3]School of Materials Science and Engineering and Electron Microscope Unit, Mark Wainwright Analytical Centre, University of New South Wales, NSW 2052, Australia

[*] These authors contributed equally to this work

[†] Corresponding to igor.aharonovich@uts.edu.au, hugo.quard@uts.edu.au

**Abstract**

*When two photonic structures of different topology are joined, an optical mode appears at their junction. Its existence depends only on this topological difference and not on the precise dimensions of either structure. Here we realize the photonic analogue of a Jackiw–Rebbi (JR) state in hexagonal boron nitride (hBN) hosting negatively charged boron vacancies ($V_B^-$), optically addressable spin defects. Two partially etched gratings, in which the order of the radiating and dark band-edge modes is reversed, are joined side by side. Angle-resolved reflectivity and photoluminescence reveal this reversal and a non-dispersive JR state at the junction. The state proves remarkably tolerant: despite fabrication errors of up to 20 nm, its wavelength lies within a 1 nm of the design, as predicted by a systematic numerical tolerance analysis. JR gratings therefore offer a robust, monolithic platform for controlling the emission of luminescent defects.*

Topological photonics transfers concepts from the topological phases of condensed matter to the control of light [1–3]. Its central prediction is the existence of optical modes at the boundary between two structures of different topology, whose presence is protected against small perturbations such as fabrication imperfections [4,5]. This robustness has enabled waveguides immune to backscattering[5,6] and lasers that operate in topologically protected modes [7–10]. It is now being brought to quantum optics, where topological modes offer a way to couple quantum emitters to nanophotonic structures that cannot be fabricated with perfect precision [11–14].

Among these boundary modes, Jackiw–Rebbi (JR) states were first described in quantum field theory as bound solutions at the interface between two regions of opposite mass [15]. A photonic analogue can be realized by joining two dielectric gratings with inverted band structures [16,17]. In each grating, the periodic modulation opens a spectral gap for the light guided in the slab, bounded by two modes of opposite symmetry: one radiates efficiently out of the plane, while the other remains confined to the slab. The two gratings are designed so that the order of these two modes is reversed from one to the other. A mode then appears at their junction, localised in space at the centre of the gap common to both gratings, and radiates as a narrow beam out of the plane [17]. Because its existence relies only on the reversed order of the two band-edge modes between the gratings, and not on their precise dimensions, the interface state is inherently tolerant of fabrication errors.

Recent experiments have used such interface states to enhance and direct the emission of excitons, either in perovskite gratings [18] or in a $WSe_2$ monolayer stacked onto $WS_2$ gratings [19]. In both cases the emitters are excitons, supplied by the grating material or by an additional layer. Hexagonal boron nitride (hBN) offers a complementary approach, whereby it itself hosts a variety of quantum emitters and luminescent spin defects [20–23]. This stable, low-loss van der Waals material can be nanostructured directly [24,25].

Here, we demonstrate a topological grating junction in hBN hosting an ensemble of boron vacancy ( $V_B^-$) defects [26]. We design two partially etched gratings of opposite topological index using full-wave simulations. The resulting interface state is described quantitatively by a non-Hermitian two-mode Hamiltonian derived from Maxwell's equations [27]. Angle-resolved reflectivity and photoluminescence reveal the band inversion of the two gratings and the JR state at their junction. The $V_B^-$ centres couple to this state and radiate through it close to the surface normal. A systematic numerical tolerance analysis shows that the state is almost insensitive to the dimensions that lithography controls least accurately.

Figure 1a illustrates the hBN topological grating investigated in this work. The structure consists of two distinct one-dimensional grating domains joined at a central interface, containing defects  within the hBN host material. The geometrical parameters of the two gratings are carefully designed such that their photonic band structures exhibit comparable bandgap widths but opposite band edge ordering. Indeed the effective Bragg-coupling coefficient $U$ characterizing the interaction between counter-propagating guided modes possesses opposite signs in the two domains. As shown below, this sign reversal creates a topological domain wall supporting a localized JR state at the interface. Emission from nearby $V_B^-$ centres can couple to this interface mode, enabling the emitted photons to be redistributed through the localized topological channel.
Each domain is a second-order grating (period $\approx \lambda/n_{eff}$), so Bragg diffraction couples the two counter-propagating guided modes of the slab both to each other and to free space. Near Γ, i.e. at normal emission, the response is non-Hermitian [27].

$$H = \omega_\Gamma \sigma_0 + v_g k \sigma_z + U\sigma_x - i\gamma_0(\sigma_0 - \sigma_x)$$

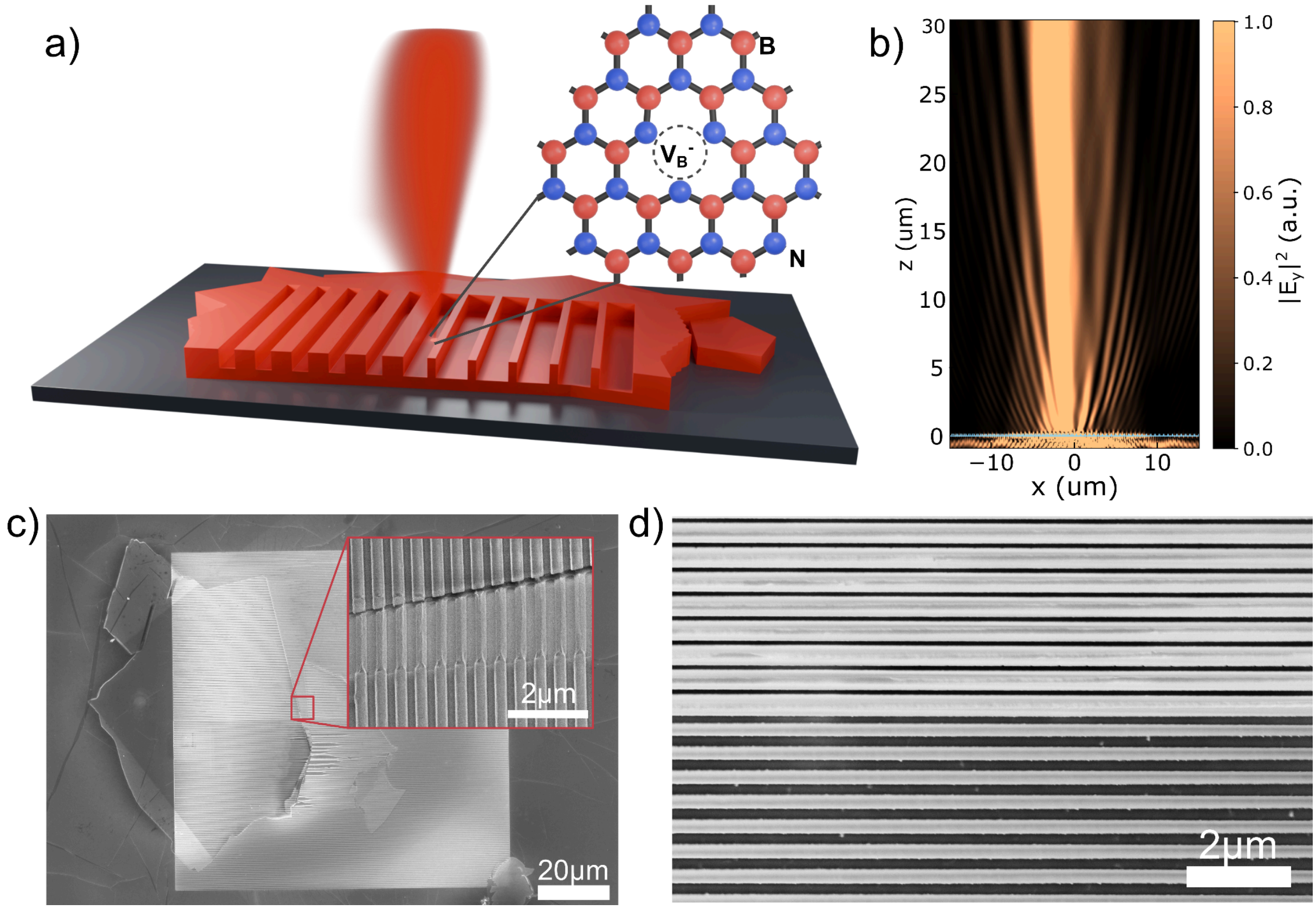

***Figure 1. Concept and emission of the topological grating.*** *a) Schematic of the device: a hBN flake on $SiO_2$/Si, partially etched into two second-order gratings with different tooth widths and periods that meet at a domain wall. $V_B^-$ centres near the junction couple to the Jackiw–Rebbi (JR) interface state bound to the wall, which radiates a narrow beam within a few degrees of the surface normal. Inset, the boron vacancy in the hBN lattice. b) Calculated intensity $|E_y|^2$ of that beam at the JR state wavelength $\lambda = 779$ nm. The near field of an in-plane dipole placed at the junction is computed by FDFD and propagated into the air above the grating by the angular-spectrum method (Methods); the grating parameters are those of Figure 2a. c) SEM image of the topological grating etched into the hBN flake, leaving part of the flake unpatterned. d) Magnified SEM image of the junction between the wide-tooth (top) and narrow-tooth (bottom) gratings. The fabricated dimensions, measured by AFM are: wide-tooth period $a_1$ = 499 nm and width $w_1$ = 280 nm; narrow-tooth period $a_2$ = 530 nm and width $w_2$ = 130 nm; etch depth $d_1$ = 103 nm; flake thickness $d_1 + d_2$ = 147 nm.*

where $\omega_\Gamma$ and $v_g$ are the frequency and group velocity of the unpatterned-slab mode at Γ, $k$ is the wavevector measured from Γ and $\sigma_i$ are the Pauli matrices. The two grating coefficients have distinct microscopic origins: the diffractive coupling $U$ is set by the second-order

Fourier component $\epsilon_2$ of the permittivity modulation and the radiative rate $\gamma_0$ by the first-order component $\epsilon_1$. The unetched hBN base is essential to this picture: it supports the guided mode even if the effective index of the narrow-tooth domain decreases and becomes closer to the underlying $SiO_2$ refractive index. It keeps the modulation perturbative, with $U$ and $\gamma_0$ small compared with $\omega_\Gamma$. At $k = 0$ the eigenvalues are

$$\omega_+ = \omega_\Gamma + U,\ \omega_- = \omega_\Gamma - U - 2i\gamma_0$$

so the bands are separated by a gap $\Delta = 2|U|$, with one edge a symmetry-protected bound state in the continuum[28,29] and the other radiatively bright. The magnitude of $U$ fixes the gap width, its sign fixes which edge lies higher and therefore acts as the topological index.
Because $U$ inherits the sign of $\epsilon_2$, it can be inverted by tuning the filling factor of the grating profile [30]. Joining two such domains therefore realizes a mass-inversion domain wall,

$$U(x) = \begin{cases} -U_0, & x < 0 \\ +U_0, & x > 0 \end{cases}$$

which maps onto the one-dimensional Jackiw–Rebbi problem with complex Dirac mass $U(x) + i\gamma_0$ (Supplementary Material). Its mid-gap solution is exponentially bound to the interface,

$$\Psi(x) \propto e^{-U_0|x|/v_g},\ \omega_{JR} = \omega_\Gamma - i\gamma_0$$

with a confinement length $L_c = v_g/U_0$ set purely by the topological coupling and a linewidth intermediate between those of the two band edges. The state exists only when $U$ changes sign across the junction. This strongly localized mode is the channel to which the emitters can couple. Because the interface state remains leaky, the light captured by this channel is redirected out of the slab as a narrow beam centred on the domain wall and emerging close to the surface normal, as shown by the calculated intensity distribution in Figure 1b.

To test this picture beyond the two-mode model, we solved Maxwell's equations for the full structure of Figure 2a, computing the dipole emission by finite-difference frequency-domain (FDFD) simulations and the reflectivity by rigorous coupled-wave analysis (RCWA; Methods). Figures 2b,d show the angle-resolved reflectivity and emission of the two gratings taken separately. Each opens a photonic bandgap at $\Delta_1 = 22.5$ nm wide for the wide-tooth grating ($w_1 = 277$ nm, $a_1 = 494$ nm, i.e. a filling factor $FF_1 = 0.56$) and $\Delta_2 = 28.0$ nm for the narrow-tooth grating ($w_2 = 110$ nm, $a_2 = 526$ nm, i.e. $FF_2 = 0.21$). The two periods compensate for the different effective indices of the two domains, so that the gaps overlap between 766 and 788.5 nm.

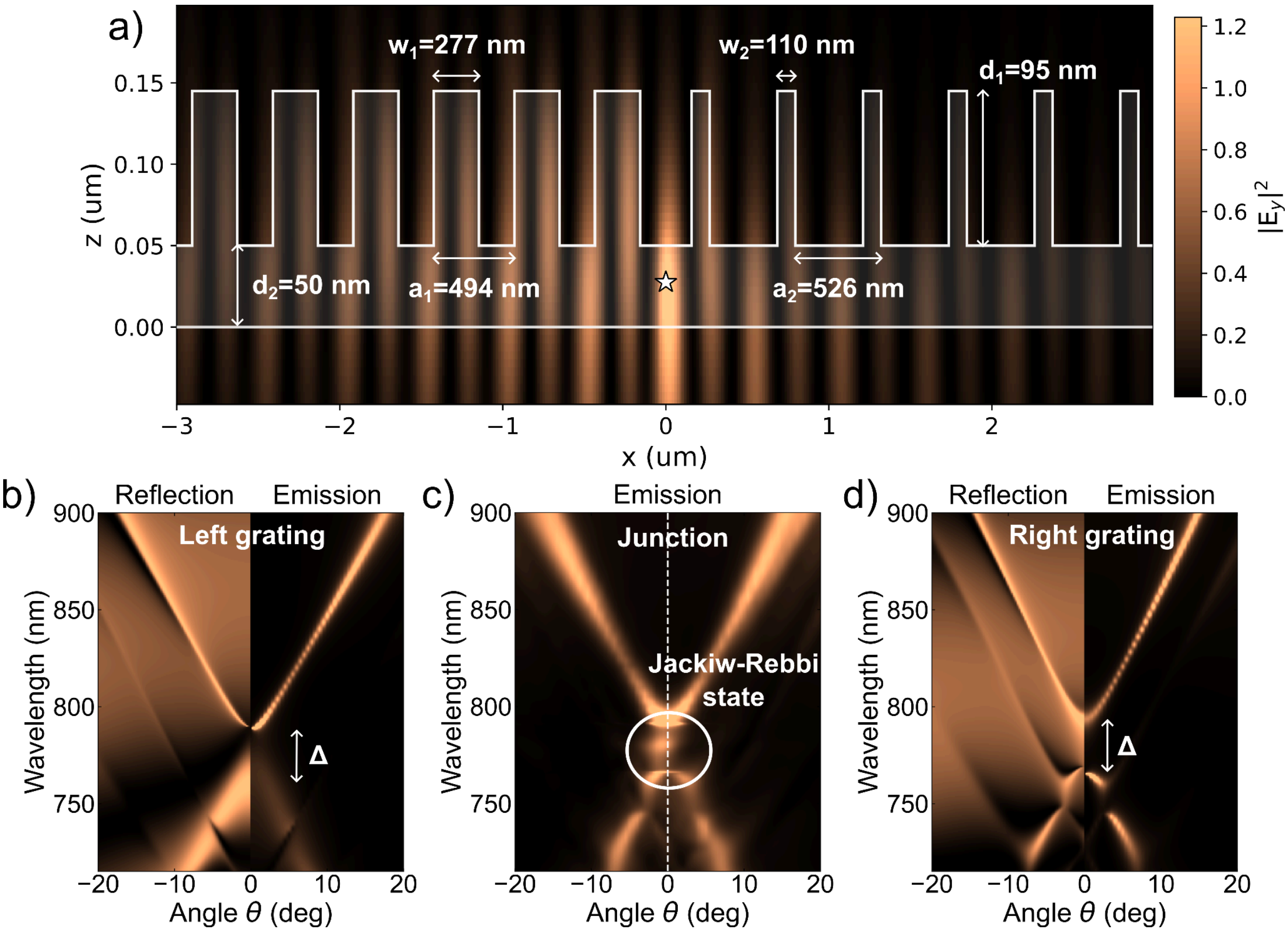

***Figure 2. Full-wave simulation of the Jackiw–Rebbi interface state.*** *a) Electric-field intensity $|E_y|^2$ at $\lambda = 779$ nm, excited by an in-plane dipole at the junction between the wide-tooth ($w_1 = 277$ nm, $a_1 = 494$ nm) and narrow-tooth ($w_2 = 110$ nm, $a_2 = 526$ nm) gratings, with the etched hBN profile overlaid ($d_1 = 95$ nm teeth on a $d_2 = 50$ nm unetched base). b,d) Angle-resolved reflectivity calculated by RCWA (left half) and dipole emission calculated by FDFD (right half) for the (b) wide-tooth and (d) narrow-tooth gratings, showing bandgaps with inverted band-edge ordering. c) Angle-resolved emission of the junction; the Jackiw–Rebbi state (circle) appears inside the common gap.*

In each grating one band edge becomes dark at normal incidence, its radiated power at Γ dropping by more than an order of magnitude, which identifies it as the symmetry-protected bound state in the continuum (BIC). The other edge is bright. The reflectivity, which probes the same modes with an external plane wave, shows the same band edges, the BIC again disappearing at normal incidence, where it cannot couple to an incident plane wave. Crucially, the ordering of the two edges is reversed between the two domains: the BIC is the long-wavelength edge (788.5 nm) of the wide-tooth grating but the short-wavelength edge (765.5 nm) of the narrow-tooth one. Because the two filling factors lie on either side of 0.5 and $\epsilon_2 \propto \sin(2\pi FF)$ (Supplementary Material), $U$ changes sign across the junction. The full-wave band structures therefore directly display the band inversion that defines the two topologically distinct phases.

When the two gratings are joined (Figure 2c), a new resonance, absent from either grating alone, appears inside the common gap at $\lambda_{JR} = 779$ nm. Unlike the band edges it is dispersionless, as expected for a state localised in real space. The near-field distribution at $\lambda_{JR}$ (Figure 2a) shows the mode pinned to the junction and decaying exponentially into both gratings.

Figure 1c,d show scanning electron microscopy images of the fabricated device. The photoluminescence (PL) of the $V_B^-$ ensemble from the unpatterned region of the flake is shown in Figure 3a. Optically detected magnetic resonance (ODMR) spectra recorded on the junction and on the unpatterned region (Figure 3b) are identical within noise, indicating that neither the fabrication nor the grating measurably alters the spin properties of the $V_B^-$ centres.

The fabricated device reproduces the calculated band structure. Figures 3c,d show the angle-resolved reflectivity and the PL of the two gratings measured separately. Both probe the same guided resonances, one through an incident plane wave, the other through the $V_B^-$ centres embedded in the flake. Each grating opens a bandgap at Γ, $\Delta = 33.0$ nm for the wide-tooth grating, between band edges at 760.2 and 793.2 nm (Figure 3c), and 46.3 nm for the narrow-tooth grating, between 749.3 and 795.6 nm (Figure 3d). In the wide-tooth grating the long-wavelength edge loses more than 70% of its amplitude within $\pm\ 0.25^{\circ}$ of normal incidence, the signature of the symmetry-protected BIC, while the short-wavelength edge stays bright. This ordering is reversed in the narrow-tooth grating: the two domains are in opposite topological phases.

Joining the two gratings produces an additional resonance inside the common gap, at 779.6 nm, absent from both gratings (Figure 3e,f). It does not disperse over the collection angle and its emission is confined to a cone with a FWHM of $1.9^{\circ}$. These are the signatures of a mode bound to the domain wall. The emission is not centred exactly at normal incidence but tilted by $0.85\ \pm\ 0.3^{\circ}$ consistent in magnitude with the $[0.5^{\circ}, 0.8^{\circ}]$ predicted by the simulations, where the tilt arises from the radiative part of the Dirac mass (Supplementary Section III).

The JR state appears in PL (Figure 3e) as well as in reflectivity (Figure 3f), so the $V_B^-$ centres couple to it and radiate through it. Because the $V_B^-$ emission band is broad and featureless, the centres act as an internal source that maps out the photonic states of the structure.

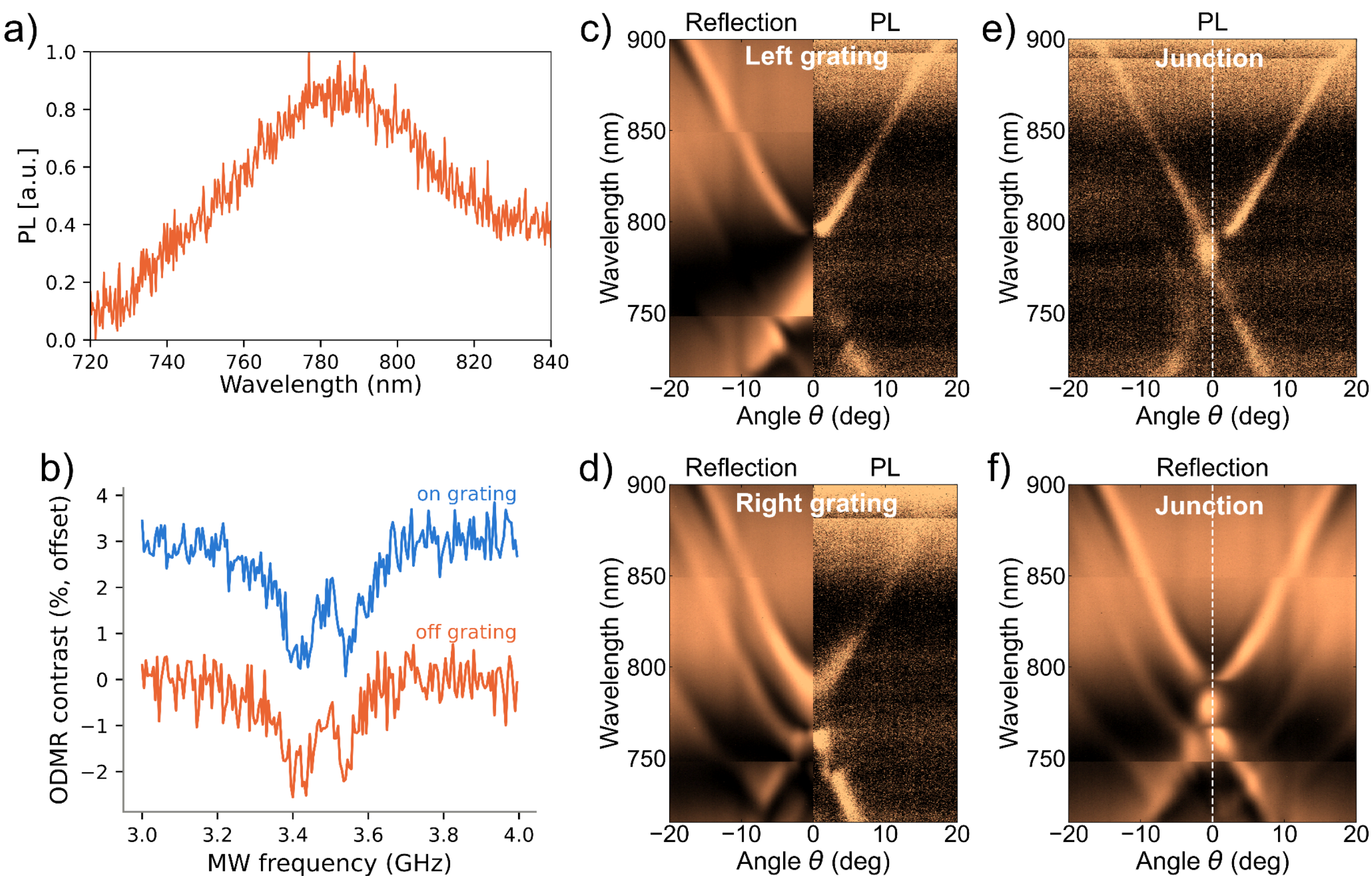

***Figure 3.*** ***Experimental demonstration of the JR interface state.*** *a) Photoluminescence of the* $V_B^-$ *ensemble measured on the unpatterned part of the flake. b) ODMR spectra recorded on the junction (blue) and on the unpatterned part of the flake (orange) with a vertical offset for clarity. c,d) Angle-resolved reflectivity (left half) and photoluminescence of the* $V_B^-$ *centres (right half) of the wide-tooth (c) and narrow-tooth (d) gratings, measured by back-focal-plane imaging (Methods). Each opens a bandgap at Γ, of 33.0 and 46.3 nm, in which one band edge vanishes at normal incidence. The dark and bright edges are interchanged between the two gratings, showing the band inversion. e,f) Photoluminescence (e) and reflectivity (f) of the junction over the full angular range. The Jackiw–Rebbi state is the non-dispersive resonance at 779.6 nm inside the common gap, emitted in a* $1.9^{\circ}$ *cone close to the surface normal (dashed line). All measurements are performed at room temperature, with the spectrometer slit perpendicular to the grooves. The horizontal discontinuities are the stitching between spectrometer grating positions.*

The measured state lies within 1 nm of the design wavelength, although the fabricated dimensions deviate from the design by up to 20 nm (Supplementary Table S3). To understand this tolerance, we repeated the junction calculation while varying each geometric parameter around its design value (Figure 4). Each map in Figure 4a–d is a set of emission spectra, the power radiated into the upper half-space, integrated over all angles, with one parameter detuned from row to row. The bright line is the JR state, and its slope gives the shift of the state per nanometer of detuning . In every sweep the state remains and moves smoothly: fabrication errors displace it in wavelength but do not remove it.

The tooth widths, which lithography controls least accurately, have almost no effect. A $\pm$ 30 nm error on either width shifts $\lambda_{JR}$ by at most 9 nm (Figure 4c,d), because the state is pinned at mid-gap whatever the magnitude of $U$ (Supplementary Section III). By contrast, the flake thickness, a common deviation in both periods, and the etch depth all shift the Bragg condition of both gratings at once. They therefore shift the whole band structure, and the state with it, without breaking the topological protection (Figure 4a,b). Figure 4e summarises all the sweeps as sensitivities, the shift of $\lambda_{JR}$ per nanometer of deviation around the design, ranked from the most to the least critical parameter. For these three parameters the sensitivities are of order unity: $+\ 1.29$ nm/nm for the thickness, $+\ 1.19$ nm/nm for the periods and $-\ 0.99$ nm/nm for the etch depth. Because thickness and etch depth act in opposite directions, a deeper etch partly compensates a thicker flake. None of these errors removes the state. It is lost only when one period is detuned from the other by more than about 15 nm, so that the two gaps no longer overlap and the state merges with a band edge. These sensitivities therefore define the accuracy that fabrication must achieve.
Applied to the dimensions measured by AFM, these sensitivities predict a net shift of only $+\ 2$ nm, placing the state at $\approx$ 782 nm, close to the measured 779.6 nm. The deeper etch almost cancels the larger periods, teeth and thickness (Supplementary Section IV).

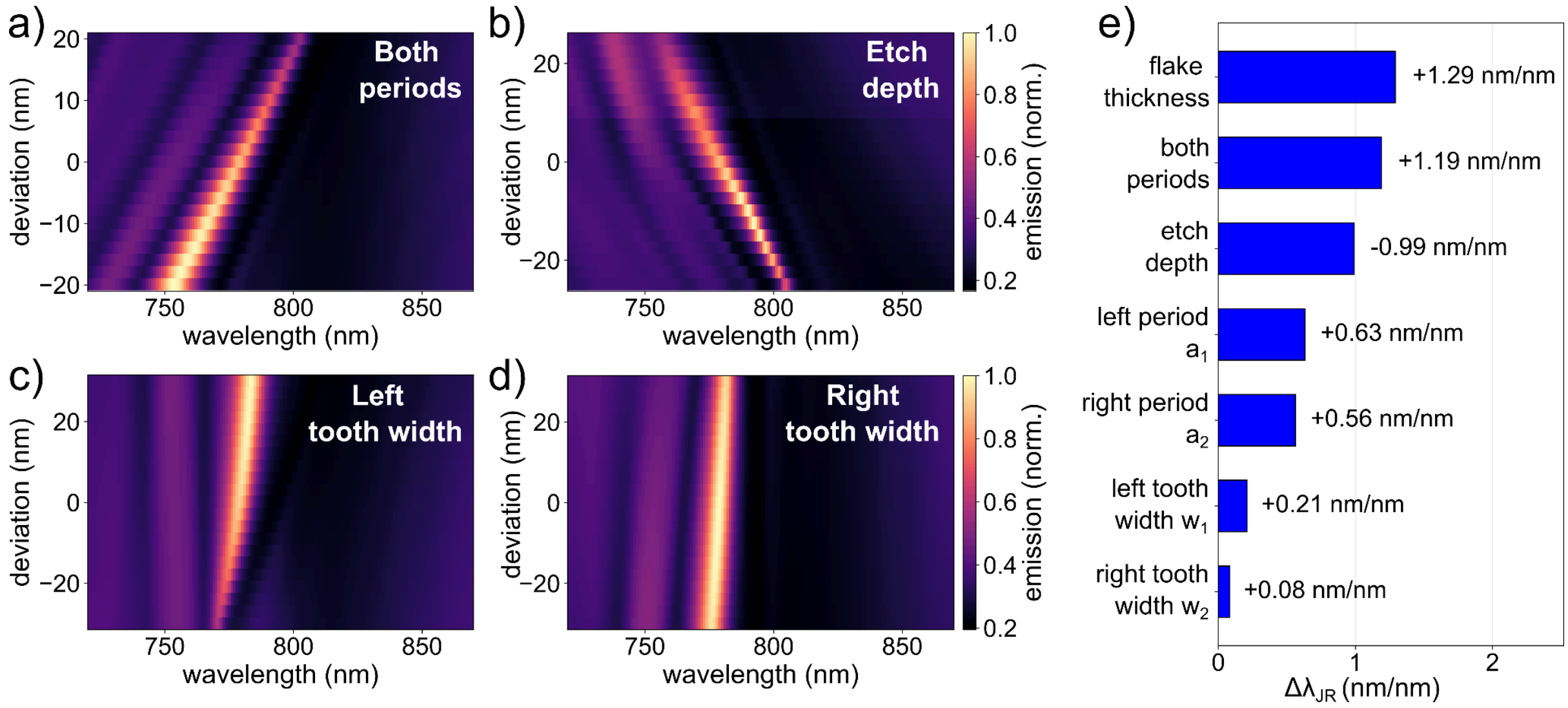

***Figure 4. Robustness of the Jackiw–Rebbi state to fabrication errors.*** *a–d) Simulated emission spectra of the junction as a function of the deviation from the design values of Figure 2a, for a common error on both periods (a), the etch depth at constant flake thickness (b), and the tooth widths* $w_1$ *(c) and* $w_2$ *(d). Each row is the power radiated into the upper half-space, integrated over all emission angles; each map is normalised to its maximum. The bright line is the Jackiw–Rebbi state. e) Sensitivity of the Jackiw–Rebbi wavelength to each geometric parameter, given as the shift of* $\lambda_{JR}$ *per nanometer of deviation and obtained from linear fits over ±10 nm around the design. The flake thickness is varied at constant etch depth.*

We have demonstrated that a domain wall between two second-order hBN gratings of opposite topological index binds a photonic JR state, and that spin defects embedded in the same flake couple to this state and radiate through it. The angle-resolved measurements reveal both elements of the argument: the dark and bright band edges are interchanged between the two gratings, which is a direct observation of the topological index, and a dispersionless resonance appears within their common bandgap at the junction. A photonic channel of this kind, robust to fabrication and compatible with the creation of the defects in situ, is therefore a promising route towards brighter spin-defect sensors in hBN.

The JR state is confined only in the direction perpendicular to the grooves. Along the grooves it extends over the entire device, so that its mode volume is determined by the patterned area rather than by the physics of the domain wall. In the future, confining the light in both directions would require a two-dimensional topological structure, in which the phase of a complex Dirac mass winds around a point rather than its sign changing along a line, as realized in Dirac-vortex cavities [31–33].

## METHODS

### Computational Modelling

### Finite-difference frequency-domain simulations.

Full-wave simulations were performed with an in-house two-dimensional finite-difference frequency-domain (FDFD) code written in Python, for the transverse-electric polarisation ($E_y$ along the grooves), which corresponds to the in-plane optical dipole of $V_B^-$ centres. The Helmholtz equation for $E_y$ is discretised with second-order central differences on a uniform grid and solved at each wavelength by direct sparse lower-upper (LU) factorisation. The domain is closed by perfectly matched layers, and the tooth edges are treated by sub-pixel permittivity averaging, which gives second-order convergence with grid spacing. hBN ($n = 2.15$, ordinary index), $SiO_2$ ($n = 1.45$) and Si ($n = 3.70$) are taken as non-dispersive. The junction, with 60 periods on each side, is excited by a point dipole at the domain wall, midway through the unetched hBN. The interface state is identified from the power confined within ±2 μm of the junction, and its quality factor from its linewidth. Angle-resolved emission is obtained by Fourier-transforming the near field above the structure ($k_x = k_0 \sin(\theta)$), and the far-field beam (Figure 1b) by propagating the same near field with the angular-spectrum method. Band structures of the individual gratings (Figure 2b,d, right halves, in the main text, and Figure S3a,c) are computed on a single unit cell with Bloch-periodic boundaries, setting $k_x = k_0 \sin(\theta)$ for each emission angle. The fabrication-tolerance analysis (Figure 4) repeats the junction calculation for each geometric variant. Full details are given in the Supplementary Methods.

**Rigorous coupled-wave analysis.**

The angle-resolved reflectivity of the individual gratings (Figure 2b,d) was calculated with the RCWA solver S4[34], on a single period with 15 Fourier harmonics under plane-wave illumination. RCWA is fast and accurate for periodic structures but ill-suited to the junction, which breaks the periodicity: the junction would have to be embedded in a periodic supercell, which adds a spurious second domain wall at each supercell boundary and requires hundreds of Fourier harmonics to resolve tens of periods, at a cost that scales as the cube of their number. The junction was therefore simulated by FDFD only.

**Fabrication**

1μm $SiO_2$/Si substrates were cleaned with acetone and isopropanol alcohol (IPA) for 15 min in an ultrasonic water bath and then blow-dried with nitrogen. Bulk undoped pristine hBN, grown using the HPHT method, was used to exfoliate hBN flakes onto the cleaned substrates using scotch tape. A combination of optical microscopy and atomic force microscopy (AFM) was used to identify flakes of the desired thickness and size.

Once the target flakes were identified, the substrates underwent a two-step spin-coating procedure. First, a positive electron-beam resist CSAR-62 (AR-P-6200.13) was spun at 5000 rpm for 60s and baked on a hot plate at 180 °C for 2 mins. Second, an anti-charging polymer, DisCharge $H_2O$, was spun at 2000 rpm to mitigate charging effects during electron beam lithography (EBL). The grating was patterned with an Elionix ELS-F125 at 125 kV with a 1 nA beam in two separate doses, one for each side of the grating, to accommodate the differing grating periods. The samples were then developed in CSAR developer 600-51, xylene and IPA for 60, 5 and 20 seconds respectively.

To remove any residual polymer in the developed regions, a brief oxygen plasma cleaning step was implemented within a Trion inductively coupled plasma reactive ion etching (ICP-RIE) using the following parameters; 12 sccm oxygen, 10W ICP, 40W RIE, under 6 mTorr of pressure for 3 seconds. To transfer the grating pattern into the hBN, a fluorine based chemistry was utilised in the same ICP-RIE with the following parameters; 60 sccm Ar, 5 sccm $SF_6$, 300W RIE under 10 mTorr of pressure. The residual CSAR was removed by submerging the sample in CSAR remover (AR-P 600-71) preheated to 135°C for one hour followed by submersion in N-Methyl-2-pyrrolidone (NMP) preheated to 90°C. The samples were then rinsed in IPA before being blow-dried using nitrogen.

A TFS Helios G5 Laser-PFIB was used to generate a focused Xe ion beam to induce knock-on collisions within the hBN lattice, thereby generating boron vacancies in the hBN grating structure. An ion dose of $1\times10^{14}$ ions/cm$^2$ was applied at an acceleration voltage of 30 kV. Scanning electron microscope (SEM) images were obtained with a Zeiss Supra 55VP.

**Angle-Resolved Reflectivity and Photoluminescence**

Angle-resolved reflectivity and photoluminescence (PL) spectra were measured with a back-focal-plane (BFP) imaging setup coupled to an imaging spectrometer. Light collected by the objective is relayed through a 4f lens system, with the first lens placed one focal length from the objective BFP. This forms a real-space image in the intermediate focal plane, where an adjustable pinhole selects the probed area of the sample: a region of $\approx$ 15 μm of diameter on a single grating, or centred on the domain wall for the junction. The second lens re-images the BFP onto the entrance slit of a custom imaging spectrometer, which consists of relay lenses, a diffraction grating and a SWIR camera. The entrance slit is aligned perpendicular to the grating grooves. It therefore transmits the emission angles θ in the plane perpendicular to the grooves, which corresponds to the Bloch wavevector $k_x = k_0 \sin(\theta)$ of the simulations, while restricting the angle along the grooves to close to zero. These angles are spectrally resolved on the camera, so that each acquisition gives the signal as a function of wavelength and θ. The spectrometer grating is mounted on a rotation stage to extend the accessible spectral range, and the wavelength calibration is adjusted for each grating position.

For reflectivity, the sample is illuminated through the objective by a collimated white-light source. For PL, the $V_B^-$ centres are excited through the same objective by a continuous-wave 405 nm laser with a power of 5.3 mW and the reflected laser light is removed by a long-pass filter at 532 nm.

**Optically detected magnetic resonance (ODMR)**

ODMR characterisation was performed using a reflection-based home-built confocal system with a 0.7 NA objective. PL was collected into a multimode fibre serving as a pinhole connected to an APD. Microwave excitation was achieved by a radio-frequency signal generator (AnaPico APSIN 4010) delivered to the sample via a suspended wire. The microwave signal was amplified using a high-power microwave amplifier (Mini-Circuits ZHL-16W-43-S+). The microwave frequency was swept within the 3.0 – 4.0 GHz resonance range. Each step consisted of 1 ms RF signal followed by a 1 ms off time to allow for a reference measurement. For each step, the ODMR contrast was calculated using PL counts as follows:

$$Contrast\ (\%) = \frac{PL(Signal) - PL(Reference)}{PL(Reference)} \times 100$$

**ACKNOWLEDGEMENT**

The authors acknowledge financial support from the Australian Research Council (CE200100010, FT220100053, DP250100973, DP260102670) and the Air Force Office of Scientific Research (FA2386-25-1-4044). The authors also acknowledge the facilities as well as the scientific and technical assistance of the Sydney Nano Foundry, Core Research Facility at the University of Sydney, part of the NSW node of NCRIS-enabled Australian National Fabrication Facility.

**COMPETING INTERESTS**

The authors declare that they have no competing interests.